\documentclass{PoS}

\def\beq{\begin{equation}}
\def\eeq{\end{equation}}
\def\bea{\begin{eqnarray}}
\def\eea{\end{eqnarray}}

\def\D0bar{\overline D{}^0}
\def\DDbar{D{}^0-\overline D{}^0}

\title{New Physics from rare decays of charm}

\ShortTitle{New Physics from rare decays of charm}

\author{\speaker{Alexey A Petrov}%
         \thanks{This work has been supported in part by the U.S. Department of Energy under contract DE-SC0007983.}\\
         Department of Physics and Astronomy\\
	Wayne State University\\
	Detroit, MI 48201, USA \\
        E-mail: \email{apetrov@wayne.edu}}

\abstract{Abundance of charm data in the current and future low energy flavor experiments makes it possible to study
rare decays of $D$-mesons with ever increased precision. I discuss theoretical implications of derived constraints on 
New Physics models from these transitions. I argue that those constraints could be competitive with results of 
direct searches for New Physics particles (including Dark Matter) at the Large Hadron Collider.}

\FullConference{16th International Conference on B-Physics at Frontier Machines\\
                 2-6 May 2016\\
                 Marseille, France}
\begin{document}

\section{Introduction}\label{Section1}

Large quantities of charmed mesons and baryons produced in high energy physics experiments make
studies of New Physics (NP) in charm transitions a natural and vibrant avenue for research. Similarly to searches for NP in 
beauty decays \cite{Nir:2016zkd}, strategies for exposition of traces of possible New Physics particles in charm transitions 
involve three main directions: (1) studies of processes that are not allowed in the Standard Model (SM), (2) studies of 
processes that are not allowed in the Standard Model at tree level, and (3) studies of  the processes that are allowed in the 
Standard Model \cite{Petrov:2010gy}. In this talk I will concentrate on the second option.

Flavor-changing neutral currents (FCNC) have been a prime vehicle of low energy NP studies in quark and 
lepton transitions for a long time. This is so primarily due to the fact that elementary currents of that type are not 
allowed in the Standard Model. It is however possible to generate such currents by quantum fluctuations, i.e. by considering
electroweak interactions at one loop level. Due to the left-handed nature of charged weak interactions in the SM, such 
currents would be induced with the coefficients that are proportional to the masses (squared) of quarks running in those loops. 
It is this fact that makes studies of NP in charm and beauty transitions very different: in $B$-physics, a huge mass of the intermediate 
top quark assures that experimentally well-studied $\Delta b = 1$ and $\Delta b = 2$ transitions are saturated by the
SM contributions. Further, the induced SM effective operators are local, which enormously simplifies theoretical calculations
of rare and $B^0-\overline B^0$ mixing transitions -- and therefore a proper interpretation of experimental data.

On the contrary, in charm, it is both relatively small mass of the intermediate bottom quark and tiny values of corresponding 
Cabbibo-Kobayashi-Maskawa (CKM) matrix elements make the short-distance SM amplitudes very small. 
This also assures that long-distance QCD effects dominate the SM predictions of most FCNC $\Delta c=1$ and $\Delta c =2$
transitions. What makes charm transitions interesting is the fact that while $\DDbar$ mixing parameters have been 
constrained \cite{Amhis:2014hma,Falk:2001hx}, rare decays of the $D^0 \to \ell^+ \ell^-$ type have never been observed. 
This fact makes them a prime target for New Physics searches in low energy experiments.

It is important to point out that decays of charmed states can probe a variety of beyond the Standard Model scenarios, with both heavy 
($m_{NP} \gg m_D$) and light ($m_{NP} \ll m_D$)  New Physics particles. 

\section{Heavy New Physics: mixing and rare decays}\label{Section2}

In general, rare decays of $D$ mesons are mediated by quark-level FCNC transitions
$c \to u \ell \ell$ and $c \to u \gamma^*$ (followed by $\gamma^* \to \ell\ell$). 
Both these decays and $\DDbar$ mixing only proceed at one loop in the SM, and, due to 
the structure of the CKM matrix, in both of these transitions Glashow-Iliopoulos-Maiani  (GIM) 
mechanism is very effective.

In this talk I will concentrate on the simplest of rare leptonic decays, $D^0 \to \ell^+ \ell^-$. This transition
has a very small SM contribution, so it could serve as a very clean probe of amplitudes. 
induced by NP particles. Other 
rare decays (such as $D \to \rho \gamma$, etc.) could receive rather significant SM contributions, which are
rather difficult to compute. For more information on those decays please see \cite{deBoer:2015boa,Fajfer:2007dy,Fajfer:2015mia,Paul:2011ar}.
There exist several experimental constraints on  $D^0 \to \ell_1^+ \ell_2^-$ transitions, resulting in the 
upper limits on flavor-diagonal and off-diagonal branching fractions \cite{Amhis:2014hma,Aaij:2013cza,Aaij:2015qmj,Petric:2010yt},
\bea
&& {\cal B} (D^0 \to \mu^+\mu^-) <  7.6\times 10^{-9}, \ \
{\cal B} (D^0 \to e^+ e^-) < 7.9\times 10^{-8}, \ \ \mbox{and}
\nonumber \\ 
&& {\cal B} (D^0 \to \mu^\pm e^\mp) <  1.3\times 10^{-8}.
\label{brs}
\eea
Theoretically, both in case of $\DDbar$ mixing and $c \to u \ell^+ \ell^-$  transitions, all possible NP contributions
can be summarized in terms of effective Hamiltonians. For the rare decays   
\beq\label{SeriesOfOperators2}
{\cal H}_{NP}^{rare}  = 
\sum_{i=1}^{10}  \frac{{\rm \widetilde C}_i (\mu)}{\Lambda^2} ~ \widetilde Q_i,
\eeq
where ${\rm \widetilde C}_i$ are the Wilson coefficients, and the $ \widetilde Q_i$ are the effective operators.
Here $\Lambda$ represents a scale of possible New Physics interactions that generate $ \widetilde Q_i$'s.
There are only ten of those operators with canonical dimension six, 
\bea
\begin{array}{l}
\widetilde Q_1 = (\overline{\ell}_L \gamma_\mu \ell_L) 
(\overline{u}_L \gamma^\mu
c_L)\ , \\
\widetilde Q_2 = (\overline{\ell}_L \gamma_\mu \ell_L)  
(\overline{u}_R \gamma^\mu
c_R)\ , \\ 
\widetilde Q_3 = (\overline{\ell}_L \ell_R) \ (\overline{u}_R c_L) \ , 
\end{array}
\qquad 
\begin{array}{l}
\widetilde Q_4 = (\overline{\ell}_R \ell_L) 
(\overline{u}_R c_L) \ , \\
\widetilde Q_5 = (\overline{\ell}_R \sigma_{\mu\nu} \ell_L) 
( \overline{u}_R \sigma^{\mu\nu} c_L)\ ,\\
\phantom{xxxxx} 
\end{array}
\label{SetOfOperatorsLL}
\eea
with five additional operators $\widetilde Q_6, \dots, \widetilde Q_{10}$ 
that can be obtained from operators in Eq.~(\ref{SetOfOperatorsLL}) by 
interchanging $L \leftrightarrow R$, e.g. $\widetilde Q_6 =  (\overline{\ell}_R \gamma_\mu \ell_R) (\overline{u}_R \gamma^\mu c_R)$,
$\widetilde Q_7 =  (\alpha/4) (\overline{\ell}_R \gamma_\mu \ell_R) (\overline{u}_L \gamma^\mu c_L)$, etc.

The Hamiltonian of Eq.~(\ref{SeriesOfOperators2}) is quite generic, so
it also contains the SM contribution usually denoted by the operators 
$Q_9 =  (\alpha/4) (\widetilde Q_1+\widetilde Q_7)$ and
$Q_{10} =  (\alpha/4) (\widetilde Q_7-\widetilde Q_1)$ (together with a substitution 
$\Lambda \to \sqrt{G_F^{-1}}$). It is worth noting that matrix elements of
several operators or their linear combinations vanish in the calculation of $ {\cal B} (D^0\to \ell^+\ell^-)$: 
$\langle \ell^+ \ell^- | \widetilde Q_5 | D^0 \rangle =
\langle \ell^+ \ell^- | \widetilde Q_{10} | D^0 \rangle = 0$ (identically),
$\langle \ell^+ \ell^- | Q_9 | D^0 \rangle \equiv  (\alpha/4) \langle \ell^+ \ell^- | (\widetilde Q_1 + \widetilde Q_7) | D^0 \rangle = 0$ 
(vector current conservation), etc.
The most general $D^0 \to \ell^+ \ell^-$ decay amplitude can be written as
\beq\label{decayampl}
{\cal M} (D^0\to \ell^+\ell^-) = {\overline u}(p_-, s_-) 
\left[ \ A + \gamma_5 B \ \right] 
v(p_+, s_+), 
\eeq
Any NP contribution described by 
the operators of Eq.~(\ref{SetOfOperatorsLL}) gives for 
the amplitudes $A$ and $B$, 
\begin{eqnarray}
\left| A\right|  &=& \frac{f_D M_D^2}{4 \Lambda^2 m_c} \left[\widetilde C_{3-8} + 
\widetilde C_{4-9}\right]\ , \nonumber 
 \\
\left| B\right|  &=& \frac{f_D}{4 \Lambda^2} \left[
2 m_\ell \left(\widetilde C_{1-2} + \widetilde C_{6-7}\right)
+  \frac{M_D^2}{m_c}
\left(\widetilde C_{4-3} + \widetilde C_{9-8}\right)
\right]\ ,  \label{DlCoeff}
\end{eqnarray}
with $\widetilde C_{i-k} \equiv \widetilde C_i-\widetilde C_k$. The amplitude of Eq.~(\ref{decayampl})
results in the branching fractions for the lepton flavor-diagonal and off-diagonal decays,
\begin{eqnarray}\label{Dllgen}
& & {\cal B} (D^0 \to \ell^+\ell^-) = 
\frac{M_D}{8 \pi \Gamma_{\rm D}} \sqrt{1-\frac{4 m_\ell^2}{M_D^2}}
\left[ \left(1-\frac{4 m_\ell^2}{M_D^2}\right)\left|A\right|^2  +
\left|B\right|^2 \right] \ \ , 
\nonumber \\
& & {\cal B} (D^0 \to \mu^+e^-) = 
\frac{M_D}{8 \pi \Gamma_{\rm D}} 
\left( 1-\frac{ m_\mu^2}{M_D^2} \right)^2 
\left[ \left|A\right|^2  + \left|B\right|^2 \right] \ \ .
\end{eqnarray}
I neglected the electron mass in the latter expression. Note that constraints on lepton flavor violating interactions, similar to 
the ones obtained from $ {\cal B} (D^0 \to \mu^+e^-)$ in Eq.~(\ref{Dllgen}), can also be obtained from two-body charmed 
quarkonium decays \cite{Hazard:2016fnc}.

According to Eq.~(\ref{DlCoeff}), the Standard Model 
contribution that appears due to $Q_9$, vanishes in the $m_\ell \to 0$ limit. 
Any NP model that contribute to $D^0 \to \ell^+ \ell^-$ can be constrained
from the bounds on the Wilson coefficients in Eq.~(\ref{DlCoeff}). It is important to
point out that because of this helicity suppression, studies of $D^0 \to e^+ e^-$ (and therefore
analyses of lepton universality in those decays) are very complicated experimentally. 
\begin{center}
\begin{table}[th]
\begin{center}
\begin{tabular}{|c|c|}
\hline 
Model & ${\cal B}(D^0 \to \mu^+\mu^-)$ \\ 
\hline
Standard Model (LD) & $\sim {\rm several} \times 10^{-13}$ \\
$Q=+2/3$ Vectorlike Singlet & $4.3 \times 10^{-11}$ \\
$Q=-1/3$ Vectorlike Singlet & $1 \times10^{-11}~(m_S/500~{\rm GeV})^2$ \\
$Q=-1/3$ Fourth Family & $1 \times 10^{-11}~(m_S/500~{\rm GeV})^2$ \\
$Z'$ Standard Model (LD) & $2.4 \times 10^{-12}/(M_{Z'}{\rm (TeV)})^2$ \\
Family Symmetry & $0.7 \times 10^{-18}$  (Case A)  \\
RPV-SUSY & $~4.8 \times 10^{-9}~(300~{\rm GeV}/m_{\tilde{d}_k})^2$ \\
Experiment & $\le 7.6 \times 10^{-9}$ \\
\hline
\end{tabular}
\end{center}
\vskip .05in\noindent
\caption{Predictions for $D^0 \to \mu^+\mu^-$ branching fraction from correlations of rare decays and $\DDbar$ mixing 
for $x_D \sim 1\%$ (from~\cite{Golowich:2009ii}). Notice that experimental constraints are beginning to probe 
charm sector of R-parity violating SUSY models.}
\label{tab:corr}
\end{table}
\end{center}
In studying NP contributions to rare decays in charm, it might be advantageous to 
study {\it correlations} of various processes, for instance
$\DDbar$ mixing and rare decays~\cite{Golowich:2009ii}. In general, one cannot predict 
the rare decay rate by knowing just the mixing rate, even if
both $x_D$ and ${\cal B}(D^0 \to \ell^+\ell^-)$ are dominated by a 
single operator contribution. It is, however, possible to do so for a restricted subset of NP 
models~\cite{Golowich:2009ii}. The results are presented in Table~\ref{tab:corr}. 

\section{Light New Physics: rare charm decays into final states with missing energy}\label{Section3}

The high-intensity $e^+e^-$ flavor factories could provide a perfect opportunity
to search for rare processes that require high purity of the final states. In particular, searches for 
$D$-decays to the final states that contain neutrinos, such as 
$D \to \pi(\rho) \nu \overline{\nu}$, are possible at those machines due to the fact that 
pairs of $D$-mesons are produced in a are charge-correlated state. Thus, there is an opportunity to tag 
the decaying heavy meson ``on the other side," which provides the charge or
CP-identification \cite{Atwood:2002ak} of the decaying ``signal" $D$ meson. This way, a variety of 
processes that are experimentally seen as transitions with ``missing energy" are possible. 

Standard Model predicts extremely small branching ratios for $D$-decay processes with 
neutrinos in the final state, i.e. ${\cal B} (D^0 \to \nu\overline\nu) \simeq 1\times 10^{-30}$, and
${\cal B} (D^0 \to \nu\overline\nu \gamma) \simeq 3\times 10^{-14}$ \cite{Badin:2010uh}. 
Thus, any detection of decays of $D$ states into channels with missing energy 
in the current round of experiments indicate 
presence of new physics. It is important to note that these NP models could be substantially
different from the models described in previous sections: experimentally, it is impossible
to say if the missing energy $\not E$ signature was generated by a neutrino or by some other 
extremely weakly-interacting particle.

Recently, a variety of models with light, $\sim {\cal O}(MeV)$ dark matter (DM) particles 
have been proposed to explain the null results of experiments designed for indirect searches 
for dark matter (see, e.g. \cite{Pospelov:2007mp,Feng:2008ya}). Such models predict couplings 
between quarks and DM particles that can be described using effective field theory (EFT) methods \cite{Petrov:2016azi}. 
These models can be tested at $e^+e^-$ flavor factories by studying $D$ (or $B$) mesons decaying 
into a pair of light dark matter particles or a pair of DM particles and a photon. The latter process 
might become important for models with fermion dark matter states as it eliminates helicity suppression of the
final state \cite{Badin:2010uh}. It is conceivable that searches for light DM in heavy meson decays 
could even be more sensitive than direct detection and other experiments, as DM couplings to 
heavy quarks could be enhanced, as happens in some models of DM, such as 
Higgs portal \cite{Pospelov:2007mp}.

Branching ratios for the heavy meson states decaying into $\chi_s \overline\chi_s$ and 
$\chi_s \overline\chi_s\gamma$, where $\chi_s$ is a DM particle of spin $s$ can be computed
in EFT framework. Since it is production of a scalar $\chi_0$
state states that is not helicity-suppressed, I only present the constraints on the models with 
scalar DM particles here. Discussion of other cases of $s=1/2$ and $s=1$ can be found in \cite{Badin:2010uh}.
A generic effective Hamiltonian for scalar DM interactions has a simple form,
\beq \label{ScalLagr} {\cal H}_{eff} =  2 \sum_i
\frac{C_i}{\Lambda^2} O_i, \eeq
where $\Lambda$ is the scale associated with the particle(s) mediating interactions between the
SM and DM fields, and $C_i$ are the Wilson coefficients. The effective operators $O_i$ are
\bea\label{ScalOper}
O_1 &=& m_c (\overline{u}_R c_L)(\chi_0^* \chi_0), \qquad
O_3 = (\overline{u}_L  \gamma^{\mu} c_L)(\chi_0^* \stackrel{\leftrightarrow}{\partial}_{\mu}\chi_0), 
\nonumber \\
O_2 &=& m_c (\overline{u}_L c_R)(\chi_0^* \chi_0), \qquad
O_4 = (\overline{u}_R \gamma^{\mu} c_R)(\chi_0^* \stackrel{\leftrightarrow}{\partial}_{\mu}\chi_0), 
\eea
where $\stackrel{\leftrightarrow}{\partial} = (\stackrel{\rightarrow}{\partial}-\stackrel{\leftarrow}{\partial})/2$ and
the DM {\it anti-particle} $\overline \chi_0$ may or may not coincide with $\chi_0$. The 
decay branching ratio for the two- body decay $D^0 \to  \chi_0 \chi_0$ is
\beq\label{Dphiphi} 
{\cal B} (D^0 \to \chi_0 \chi_0) = 
\frac{\left(C_1-C_2\right)^2}{4\pi M_{D} \Gamma_{D^0}}
\left[\frac{f_{D}
M_{D}^2 m_c}{\Lambda^2 (m_c +m_q)}\right]^2
\sqrt{1-4 x_\chi^2} 
 \eeq
where $x_\chi  = {m_\chi}/{M_{D^0}}$ is a rescaled DM
mass. Clearly, this rate is not helicity-suppressed, so it could be
quite a sensitive tool to determine DM properties at $e^+e^-$ flavor factories. 

Applying the formalism described above, distribution of the photon energy and decay width of the 
radiative transition $D^0 \to \chi_0\chi_0\gamma$ can be computed,
\begin{eqnarray}
\label{dGammaDphiphigamma}
\frac{d\Gamma}{dE_{\gamma}}(D^0 \to \chi_0 \chi_0\gamma)  &=& \frac{f_{D}^2
\alpha C_3 C_4}{3 \Lambda^4}\left(\frac{F_{D}}{4\pi}\right)^2
\frac{2M_{D}^2 E_\gamma(M_{D}(1-4x_\chi^2)-2E_\gamma)^{3/2}}{\sqrt{M_{D}-2E_\gamma}}\\
\label{GammaDphiphigamma} 
{\cal B} (D^0 \to \chi_0 \chi_0\gamma)&=& \frac{f_{D}^2 \alpha
C_3 C_4 M_{D}^5}{6 \Lambda^4 \Gamma_{D^0}} 
\left(\frac{F_{D}}{4\pi}\right)^2 \\
&\times&  \left(\frac{1}{6}\sqrt{1-4x_\chi^2}(1 - 16 x_\chi^2 - 12x_\chi^4)
-12x_\chi^4\log\frac{2 x_\chi}{1+\sqrt{1-4x_\chi^2}} \right),
\nonumber
\end{eqnarray}
We observe that Eqs.~(\ref{dGammaDphiphigamma}) and (\ref{GammaDphiphigamma}) do
not depend on $C_{1,2}$. This can be most easily seen from the
fact that $D \to \gamma$ form factors of scalar and pseudoscalar
currents are zero. This implies that studies of both $D^0 \to \not E$
and $D^0 \to \gamma \not E$ processes probe complementary operators in the effective 
Hamiltonian of Eq.~(\ref{ScalLagr}). Similar conclusions follow for the decays of 
beauty-flavored mesons into the final states with missing energy,
where energy scales (DM heavy mediator masses) of order 10 TeV are 
probed by currently available data \cite{Badin:2010uh}.

\section{Probing rare charm transitions in production experiments}\label{Section4}

As was mentioned in Section \ref{Section3}, studies of lepton universality in 
$D^0 \to \ell^+ \ell^-$ decays could be complicated due to helicity suppression of this 
decay, which makes the branching ratio of $D^0 \to e^+ e^-$ tiny. This feature 
persists in many models of NP. An interesting alternative to $D^0 \to e^+e^-$ process that 
is {\it not} helicity-suppressed is a related decay $D^{*}(2007)^0 \to e^+e^-$.
While also probing the FCNC $c\bar{u} \to \ell^+\ell^-$ transition, this decay 
is sensitive to the contributions of operators that $D^0 \to \ell^+\ell^-$ cannot be sensitive to. 
Unfortunately, a direct study of the $D^* \to e^+e^-$  decay is practically impossible, 
since the $D^*$  decays strongly or electromagnetically.  

\begin{figure}[t]
\begin{center}
\includegraphics[width=6cm]{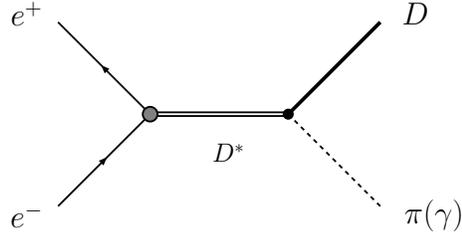}
\end{center}
\caption{\label{figure} Probing the $ c\bar{u}\to e^+ e^-$ vertex
with the $D^*(2007)^0$ resonance production in $e^+e^-$ collisions.}
\end{figure}

Nevertheless, it might be possible to probe the $D^*\to e^+e^-$ transition experimentally \cite{Khodjamirian:2015dda}. 
Assuming time-reversal invariance, it would be equivalent to measure the corresponding {\it production} process 
$e^+e^- \to D^*$, as shown in Fig.~\ref{figure}. In order to do so, a run of an $e^+e^-$ collider, such as BEPCII or VEPP-2000, 
at the center-of-mass energy corresponding to the mass of the $D^*$ meson, $\sqrt{s} \approx 2007$ MeV, should be performed. 
If produced, the $D^{*0}$ resonance will decay via strong ($D^{*0}\to D^0\pi^0$) or electromagnetic ($D^{*0}\to D^0\gamma$)  interactions  
with branching fractions of $(61.9\pm 2.9)\%$ and 
$(38.1\pm 2.9)\%$ respectively.\footnote{Note that the charged mode 
$D^{*0} \to D^+\pi^-$ is forbidden by the lack of the available phase space.}
A single charmed particle in the final state at this $\sqrt{s}$ could serve as an excellent tag for such 
process, with other sources of production of a single $D$ meson being negligibly small \cite{Khodjamirian:2015dda}.
A thorough study of both short-distance and long-distance contributions to $e^+e^- \to D^*$ in the Standard Model and in 
NP models has been performed. This process, albeit very rare, has clear advantages with respect to the
 $D \to e^+e^-$ decay: the helicity suppression is absent, and a richer set of effective operators can be probed. 
Employing the most recent values of the Wilson coefficients for $c\to u$ transition \cite{deBoer:2016dcg}, it
was shown that, contrary to other rare decays of charmed mesons, long-distance SM contributions are under 
theoretical control and contribute at the same order of magnitude as the short-distance ones. Similar opportunities exist for 
$B$-decays as well \cite{Khodjamirian:2015dda,Grinstein:2015aua}.

\section{Conclusions and outlook}\label{Section5}

The absence of any hints of new particles from direct searches at the LHC experiments makes careful 
studies of their possible quantum effects an important tool in our arsenal of methods for probing physics
beyond the Standard Model. Abundance of charm data in the current and future low energy flavor experiments makes 
it possible to study New Physics in rare decays of $D$-mesons with ever increased precision. The obtained constraints 
from a variety of methods described in this talk could be competitive with results of continuing direct searches for New Physics 
particles at the Large Hadron Collider.


\end{document}